\documentclass[prb,twocolumn,superscriptaddress,floatfix]{revtex4}
\usepackage{graphicx}
\usepackage{color}
\usepackage{hyperref}
\usepackage{multirow}

\newcommand{\cm}{\mbox{cm}\ensuremath{^{-1}}}

\newcommand{\dgC}{\ensuremath{^\circ\mbox{C}}}

\begin{document}
\title{Electromagnon in the Z-type hexaferrite $({\rm Ba}_{x}{\rm Sr}_{1-x})_3\rm Co_2Fe_{24}O_{41}$}

\author{Filip Kadlec}
\affiliation{Institute of Physics, Czech Academy of Sciences, Na Slovance 2, 182 21 Prague~8, Czech Republic}
\author{Christelle Kadlec}
\affiliation{Institute of Physics, Czech Academy of Sciences, Na Slovance 2, 182 21 Prague~8, Czech Republic}
\author{Jakub V\'\i t}
\affiliation{Institute of Physics, Czech Academy of Sciences, Na Slovance 2, 182 21 Prague~8, Czech Republic}
\affiliation{Faculty of Nuclear Science and Physical Engineering, Czech Technical University, B\v{r}ehov\'{a} 7, 115 19 Prague~1, Czech Republic}
\author{Fedir Borodavka}
\affiliation{Institute of Physics, Czech Academy of Sciences, Na Slovance 2, 182 21 Prague~8, Czech Republic}
\author{Martin Kempa}
\affiliation{Institute of Physics, Czech Academy of Sciences, Na Slovance 2, 182 21 Prague~8, Czech Republic}
\author{Jan Prokle\v{s}ka}
\affiliation{Department of Condensed Matter Physics, Faculty of Mathematics and Physics, Charles University, Ke Karlovu 5, 121 16 Prague 2, Czech Republic}
\author{Josef Bur\v{s}\'\i k}
\affiliation{Institute of Inorganic Chemistry, Czech Academy of Sciences, 250 68 \v{R}e\v{z}, Czech Republic}
\author{R\'obert Uhreck\'y}
\affiliation{Institute of Inorganic Chemistry, Czech Academy of Sciences, 250 68 \v{R}e\v{z}, Czech Republic}
\author{St\'ephane Rols}
\affiliation{Institut Laue-Langevin, Bo\^\i te Postale 156, 38042 Grenoble Cedex 9, France}
\author{Yi Sheng Chai}
\affiliation{Institute of Physics, Chinese Academy of Sciences, Beijing, P. R. China}
\author{Kun Zhai}
\affiliation{Institute of Physics, Chinese Academy of Sciences, Beijing, P. R. China}
\author{Young Sun}
\affiliation{Institute of Physics, Chinese Academy of Sciences, Beijing, P. R. China}
\author{Jan Drahokoupil}
\affiliation{Institute of Physics, Czech Academy of Sciences, Na Slovance 2, 182 21 Prague~8, Czech Republic}
\author{Veronica Goian}
\affiliation{Institute of Physics, Czech Academy of Sciences, Na Slovance 2, 182 21 Prague~8, Czech Republic}
\author{Stanislav Kamba}
\thanks{Authors to whom correspondence should be addressed}
\email[e-mail: ]{ kamba@fzu.cz; kadlecf@fzu.cz}
\affiliation{Institute of Physics, Czech Academy of Sciences, Na Slovance 2, 182 21 Prague~8, Czech Republic}

\begin{abstract}
We studied experimentally the high-temperature  magnetoelectric
$({\rm Ba}_{x}{\rm Sr}_{1-x})_3\rm Co_2Fe_{24}O_{41}$ prepared as  ceramics
($x=0$, 0.2) and a single crystal ($x=0.5$)
using inelastic neutron scattering, THz time-domain, Raman and far-infrared
spectroscopies. The spectra, measured with varying temperature and
magnetic field, reveal rich information about the collective spin and lattice
excitations. In the ceramics, we observed an infrared-active
magnon which is absent in \textbf{E}$^{\omega}\perp z$ polarized THz
spectra of the crystal,  and we assume that it is
an electromagnon active in \textbf{E}$^{\omega} \| z$  polarized spectra.
On heating from 7 to 250\,K, the frequency of this electromagnon drops
from 36 to 25\,\cm{} and its damping gradually increases, so it becomes overdamped at room temperature.
Applying external magnetic field has a similar effect on the
damping and frequency of the electromagnon, and the mode
is no more observable in the THz spectra above 2\,T, as the transverse-conical
magnetic structure transforms into a collinear one. Raman spectra reveal
another spin excitation with a slightly different frequency and much higher damping.
	Upon applying magnetic field higher than 3\,T,  in the low-frequency part of the THz
		spectra, a narrow
		excitation appears whose frequency linearly increases with magnetic field. We interpret this feature as the ferromagnetic resonance.
\end{abstract}
\date{\today}

\maketitle \section{Introduction}

The current research on multiferroic materials is motivated by both
an incomplete understanding of their fundamental physical properties and their potential in
realizing novel devices which would make use of static or dynamic
magnetoelectric (ME) coupling in non-volatile memories, spintronics, magnonics
and microwave devices. For practical applications, a strong ME coupling
near room temperature is required. BiFeO$_3$, the most intensively
studied multiferroic, exhibits a large
ferroelectric (FE)
polarization and multiferroicity up to 643\,K. However, its spiral
magnetic structure persists up to an external magnetic
field of 19\,T and, therefore, the ME coupling is rather low up to
18\,T.\cite{Tokunaga10} Much larger ME coupling was observed in
spin-order-induced ferroelectrics.\cite{Tokura14-review} Unfortunately, most of these materials
exhibit multiferroic properties only below ca.\ 50--100\,K.
Nevertheless, some ferrites with hexagonal symmetry, called hexaferrites,
exhibit magnetic-field-induced electric polarization close to or even far
above room temperature and their ME coupling can be very
high.\cite{Chun12,Chai14,Kimura12}

Based on their chemical formulas and crystal structures,
hexaferrites can be classified into several types: \textit{M}-type, such as
$\rm(Ba,Sr)Co_{\it x}Ti_{\it x}Fe_{12-2{\it x}}O_{19}$, \textit{Y}-type
$\rm(Ba,Sr)_2Me_2Fe_{12}O_{22}$, \textit{W}-type $\rm(Ba,Sr)Me_2Fe_{16}O_{27}$,
\textit{Z}-type $\rm(Ba,Sr)_3Me_2Fe_{24}O_{41}$, \textit{X}-type
$\rm(Ba,Sr)_2Me_2Fe_{28}O_{46}$, and \textit{U}-type
$\rm(Ba,Sr)_4Me_2Fe_{36}O_{60}$, where Me is a bivalent metal ion (e.g.\ Co, Mg,
Zn).\cite{Kimura12,Pullar12} The structures of hexaferrites can be described as
sequences of three basic building blocks
(usually denoted by \textit{S, R}, and \textit{T}) periodically stacked  along the \textit{z}
axes. Since the hexagonal structures of hexaferrites are associated
with centrosymmetric $P6_{3}/mmc$ or $R \bar 3m$ space groups, no FE
polarization should exist in these materials. Nevertheless, their
ferrimagnetic structures are spiral or heliconical and they can be
easily changed by external magnetic field to transverse conical, where an
electric polarization of the order of tens of $\rm \mu C/m^2$ can appear
due to the inverse Dzyaloshinskii-Moriya interaction ($\propto \textbf{S}_{i}\times\textbf{S}_{j}$) between non-parallel
spins $ \textbf{S}_{i}$ and $\textbf{S}_{j}$. In this case the centrosymmetric structure  should be broken and the
crystal symmetry  lowered. Importantly,  the  magnetic field needed to induce a
polarization can be very low (of the order of millitesla) and the effect can be remanent.\cite{Chai14, Shen14}
By contrast, the polarization
disappears in higher magnetic fields (usually above 2\,T) when the magnetic structure changes.\cite{Kimura12}

ME properties of the Z-type hexaferrite $\rm
Sr_3Co_2Fe_{24}O_{41}$ were reported first time by Kitagawa \textit{et
al.}\cite{Kitagawa10} They discovered that this highly resistive material  exhibits magnetic-field induced electric polarization  at least up to 300\,K. ME and magnetodielectric
effects in $\rm Sr_3Co_2Fe_{24}O_{41}$ were confirmed later by
other authors.\cite{Soda11,Wu12,Zhang12,Ebnabbasi12a,Ebnabbasi12b,Okumura13}
Chun \textit{et al.}\cite{Chun12} investigated $\rm(Ba_{{\it x}}Sr_{1-{\it
x}})_3Co_2Fe_{24}O_{41}$ and discovered that the ME effect is the highest in
$\rm(Ba_{0.52}Sr_{2.48})Co_2Fe_{24}O_{41}$. Moreover, for this
composition, the magnetic structure changes from transverse conical to
collinear ferrimagnetic structure at a temperature as high as 400\,K, so the
ME effect  can be measured well above room
temperature.\cite{Soda11,Chun12}

In general, depending on the crystal and magnetic structures, the ME
coupling  may be due to one of three different mechanisms:
exchange striction (magnetostriction $\propto \textbf{S}_{i}\cdot\textbf{S}_{j}$), inverse Dzyaloshinskii-Moriya interaction
or spin-dependent hybridization of the \textit{p} and \textit{d}
orbitals.\cite{Tokura14-review} The same mechanisms can be also responsible for
the dynamic ME coupling, which activates magnons in the THz
or far-infrared \textit{dielectric permittivity} spectra and therefore
they are called \textit{electromagnons}.\cite{Pimenov06,Kida09,Tokura14-review}
By contrast, common magnons impact only
upon the magnetic susceptibility spectra in the microwave or THz
ranges. They are also called ferromagnetic (FMR) and
antiferromagnetic resonances, and their frequencies
correspond to acoustic-like and optic-like magnons, respectively, with
wavevectors $q$ from the Brillouin-zone center ($q = 0$). As for
electromagnons, they have frequently wavevectors  out of the
Brillouin-zone center ($q\neq0$) and they can be excited by THz
photons (with a wavevector $q \sim 0$) only if the magnetic structure is modulated, which
is true in practically all spin-order-induced multiferroics. As regards
the hexaferrites, an electromagnon was reported only in the
$Y$-type compound $\rm Ba_2Mg_2Fe_{12}O_{22}$.\cite{Kida09a,Kida11,Nakajima16}
Interestingly, it was observed not only in the spin-induced FE phase
below 50\,K, but also in the paraelectric one at 90\,K, if an
external magnetic field ($0.4\,{\rm T}\leq \mu_0 H \leq 1.6\,{\rm T}$) was
applied ; in that case the magnetic structure
changed from proper screw to longitudinal (for $\textbf{\textit{H}}
\parallel z$) or transverse  conical (for $\textbf{\textit{H}} \perp z$).\cite{Kida11}
The  activation of the electromagnon in \textbf{E}$^{\omega}\parallel z$ polarized spectra was explained by the
exchange striction,  although the static electric
polarization \textbf{P}$ \perp z$ in hexaferrites comes from the inverse Dzyaloshinskii-Moriya
interaction.\cite{Kida11}

Since the $Z$-type hexaferrites $({\rm Ba}_{x}{\rm Sr}_{1-x})_3\rm
Co_2Fe_{24}O_{41}$ exhibit ferrimagnetic phase transitions at temperatures as
high as $T_{\rm N}=$ 700\,K and ME coupling  up to nearly 400\,K,
one can expect electromagnons to be activated in their THz spectra at much
higher temperatures than in $Y$-type hexaferrites. For that reason we have
performed detailed THz time-domain, infrared (IR), Raman and inelastic neutron
scattering (INS) spectroscopic studies from 5 to 900\,K on  $({\rm Ba}_{x}{\rm
Sr}_{1-x})_3\rm Co_2Fe_{24}O_{41}$ ceramics with $x=0$ and 0.2 and a single
crystal with $x=0.5$. Three spin excitations including one electromagnon were discovered.

\section{Samples and experimental details}

Powders of hexagonal ferrite with a composition $({\rm Ba}_{x}{\rm
Sr}_{1-x})_3\rm Co_2Fe_{24}O_{41}$  ($x=0$ and 0.2) were prepared by the Pechini
type in-situ polymerizable complex method relying on immobilization of
metalloorganic precursor complexes in a rigid organic polymer network,
thus ensuring the compositional homogeneity of the complex oxide. First,
calculated amounts  of strontium carbonate (SrCO$_3$),
barium carbonate (BaCO$_3$), cobalt nitrate (Co(NO$_3$)$_2\cdot
6$H$_2$O), and iron nitrate (Fe(NO$_3$)$_3\cdot 9$H$_2$O; all
chemicals from
Sigma-Aldrich) were decomposed in a 0.1 mol/l solution of nitric acid in
distilled water. After their complete dissolution, a calculated amount of
a polymer gel formed by reaction of citric acid
(HOOCCH$_2$C(OH)-(COOH)CH$_2$COOH) with ethylene glycol
(HOCH$_2$CH$_2$OH) in water was added to this solution, mixed and
heated to 130\dgC. With continued heating over several hours the clear solution
became highly viscous, gradually gelled and finally polymerized into a
voluminous resin. After breaking the resin, its drying (at 150\dgC) and charring
(at 350\dgC) for 24\ hrs, the resulting powder was heat-treated in
an oxygen atmosphere at 1200\dgC\, for 12\ hrs. A powder X-ray diffraction measurement
proved a single-phase composition of the product. Cold
isostatic pressing (pressure 300\,MPa) and subsequent sintering at
1200\dgC{} in
oxygen atmosphere were used to obtain dense ceramics of the Z-phase
ferrite. Single- and double-side polished pellets with diameters
of 6\,mm and thicknesses of 2 and 0.48\,mm were prepared for the IR and THz
studies, respectively.

A $\rm(Ba_{0.5}Sr_{0.5})_3Co_2Fe_{24}O_{41}$ single crystal with a natural
hexagonal plane was grown by the flux method.\cite{Momozawa87} It had a diameter
of 4--5\,mm and a thickness of 1.8\,mm.

Low-temperature IR  reflectivity measurements in the frequency range
30--670\,\cm (or, equivalently, 1--20\,THz) were performed using a Bruker
IFS-113v Fourier-transform IR spectrometer equipped with a liquid-He-cooled Si
bolometer (1.6\,K) serving as a detector. Room-temperature mid-IR
spectra up to 5000\,\cm{} were obtained using a pyroelectric deuterated triglycine sulfate detector.

THz complex transmittance from 3
to 50\,\cm (with the upper limit due to sample opacity) was measured using a
custom-made time-domain spectrometer. For the low-temperature IR reflectivity and THz complex transmittance spectroscopy, Oxford Instruments Optistat cryostats with mylar and polyethylene windows, respectively, were used.  THz spectroscopy with magnetic field was performed using a custom-made time-domain spectrometer comprising an Oxford Instruments Spectromag cryostat with a superconducting magnet, allowing us to apply an external magnetic field of up to 7\,T; the Faraday geometry (wavevector parallel to the magnetic field) was used.

INS was measured on a powder sample
(9.75\,g) using the neutron time-of-flight spectrometer IN4 in the Institut
Laue-Langevin in Grenoble, France.

For Raman studies, a Renishaw RM\,1000 Micro-Raman spectrometer equipped with a
CCD detector and Bragg filters was used. The experiments were performed in
the backscattering geometry within the 10--800\,\cm\ range using an
Ar laser with the wavelength of 514\,nm and an Oxford Instruments Optistat optical continuous He-flow cryostat.
Further, using a Quantum design PPMS 9T
	instrument, we carried out measurements of the magnetic susceptibility,
magnetization, and of the ME effect, in a temperature
interval from 5 to 1000\,K with a magnetic field of up to 9\,T.

\section{Results and discussion}

\subsection{Magnetic and magnetoelectric properties}

\begin{figure} \centering \includegraphics[width=0.9\columnwidth]{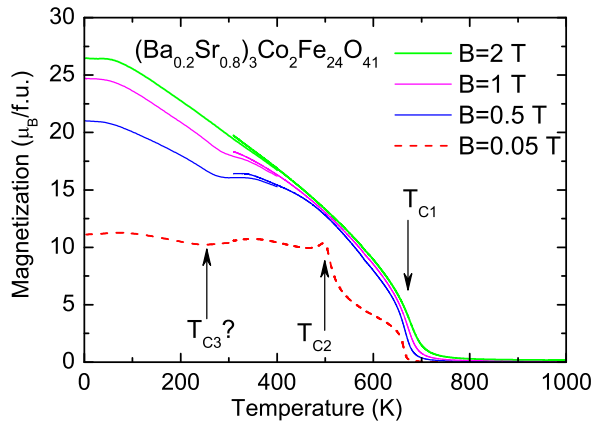}
	\caption{Temperature dependence of the magnetization for
	polycrystalline $\rm(Ba_{0.2}Sr_{0.8})_3Co_2Fe_{24}O_{41}$ taken at
	various magnetic fields. The temperatures of
	magnetic phase transitions are
	marked by arrows.} \label{fig:ZBaM-T} \end{figure}

Fig.~\ref{fig:ZBaM-T} shows the magnetization $M$ of the
$\rm(Ba_{0.2}Sr_{0.8})_3Co_2Fe_{24}O_{41}$ ceramics as a function of temperature
for several values of magnetic field. These dependences are similar to
the previously published  data obtained with $\rm
Sr_{3}Co_2Fe_{24}O_{41}$ ceramics\cite{Kitagawa10} and a $\rm
Ba_{0.52}Sr_{2.48}Co_2Fe_{24}O_{41}$ single crystal.\cite{Chun12} Thus,
the magnetic phase diagram of (${\rm Ba}_{x}{\rm Sr}_{1-x})_3\rm
Co_2Fe_{24}O_{41}$ is probably relatively independent of $x$ at least for $0\leq x \leq0.2$. The collinear
ferrimagnetic structure with spins parallel to the $z$ axis appears below
	$T_{\rm C1} \approx 700\,\rm K$. At $T_{\rm C2} \approx 500\,\rm K$, the
	spins start to rotate towards the $xy$-plane and at $T_{\rm con} \approx 400\,\rm K$ a transverse
conical structure is stabilized.\cite{Chun12} Near 260--300\,K, we observed shallow minima
	in the $M(T)$ curves, similarly to Refs.~\onlinecite{Soda11,Chun12}.
This feature has not been satisfactorily explained yet, and we suppose
it is due to some further changes in the magnetic structure.  Subsequently,
we measured magnetization curves $M(H)$ at various
temperatures (see Fig.~1 in Suppl.\ Materials\cite{Suppl-hexaferrite}) and found some
features due to magnetic phase transitions in the range from 0.1 to 1\,T.
The magnetization reversal processes during these transitions are explained in Ref.~\onlinecite{Tang16}. ME measurements of $P(H)$ at 10\,K (see Fig.~2 in Suppl.\ Materials\cite{Suppl-hexaferrite}) revealed changes in electric polarization induced by magnetic field greater than 0.01\,T, attaining a maximum at 0.3\,T and vanishing near 2\,T. A similar behavior  was reported for a single
crystal\cite{Chun12}, where the maximum polarization was, two orders of magnitude higher than in our ceramics. For temperatures higher than 50\,K, a strong leakage conductivity prevented us from
acquiring meaningful ME data.

\subsection{Phonon spectra }

We performed the factor-group analysis of lattice vibrations in
(Ba,Sr)$_3$Co$_2$Fe$_{24}$O$_{41}$ for the centrosymmetric hexagonal space
group $P6_{3}/mmc$ ($D_{6h}^{4}$), taking into account
the site
symmetries published in Refs. \onlinecite{Takada03,Takada06} with the same
crystallographic sites shared between Fe/Co and Ba/Sr atoms. The unit
cell contains $2\times70$ atoms, and the analysis yields:
\begin{eqnarray}
\Gamma_{D_{6h}^{4}}&=29A_{2u}(z)+37E_{1u}(x,y)+26A_{1g}(x^2+y^2,z^2)+\nonumber \\
&+33E_{1g}(xz,yz)+36E_{2g}(x^2-y^2,xy)+27B_{2u}+\nonumber \\
&+34E_{2u}+28B_{1g}+7A_{1u}+8A_{2g}+8B_{1u}+7B_{2g} \nonumber\\
\end{eqnarray}
where $x$, $y$, and $z$ mark electric polarizations of the IR radiation for
which the phonons are IR active, whereas the rest of the symbols in
parentheses are components of the Raman tensor. After subtracting two acoustic
phonons, 64 IR-active and 95 Raman-active phonons are predicted in the spectra.
Additional 119 phonons are silent, i.e.\ inactive in the IR or Raman
spectra.

Assuming that the crystal structure undergoes an
equi-translational spin-induced FE phase transition, the soft
mode in the paraelectric phase should have, according to the tables
in Ref.~\onlinecite{Janovec75}, the $A_{2u}$ symmetry, and the resulting
acentric space group will be $P6mm$ ($C_{6v}^{1}$). The factor-group analysis
of phonons from the Brillouin-zone center reads
\begin{eqnarray}
\Gamma_{C_{6v}^{1}}&=55A_{1}(z,x^2+y^2,z^2)+70E_{1}(x,y,xz,yz)+55B_{2}\nonumber \\
&+70E_{2}(x^2-y^2,xy)+15A_{2}+15B_{1}\ .
\end{eqnarray}
Thus, in the FE phase, one can expect 123 IR-active modes, 193 modes can be theoretically observed in Raman spectra and 85 modes remain silent. Nevertheless, the intensities of the newly activated modes are expected to be
very low, because the polar distortion (proportional to the
polarization) in spin-order-induced ferroelectrics and probably also in (${\rm Ba}_{x}{\rm Sr}_{1-x})_3\rm Co_2Fe_{24}O_{41}$ is four
orders of magnitude smaller than in the canonical FE BaTiO$_3$. We
note that, due to this fact, no new phonons below $T_{\rm C}$ were reported also
in other spin-induced ferroelectrics, only small shifts of phonon frequencies
were observed due to the spin-phonon coupling.\cite{Schmidt09,Moeller14}
Anyway, the factor-group analysis is useful for the
discussion of electromagnon activity in both IR and Raman spectra below.

\begin{figure}[]
	  \centering
	  \includegraphics[width=88mm]{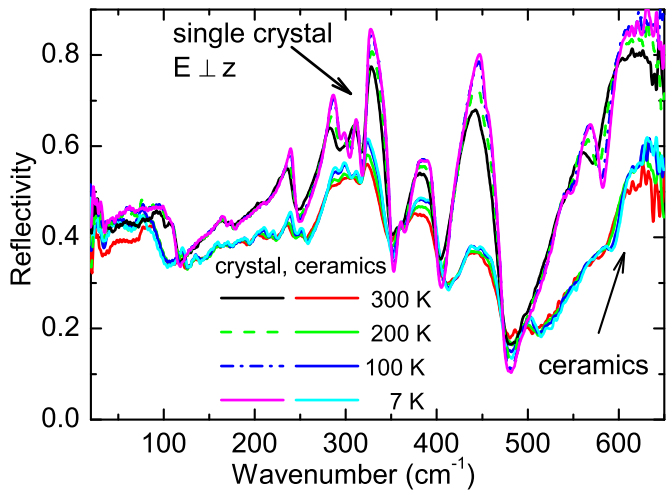}
	  \caption{Temperature dependence of IR reflectivity spectra of
		  $\rm(Ba_{0.2}Sr_{0.8})_3Co_2Fe_{24}O_{41}$ ceramics and a $\rm (Ba_{0.5}Sr_{0.5})_3Co_2Fe_{24}O_{41}$ single crystal. The latter spectrum was taken in polarization \textbf{E}$^{\omega}\perp z$ and \textbf{H}$^{\omega}\perp z$, i.e. only $E_{1u}$ symmetry phonons are seen.}
	  \label{fig:IRrefl}
  \end{figure}

Fig.\,\ref{fig:IRrefl} shows the IR reflectivity spectra of the
$\rm(Ba_{0.2}Sr_{0.8})_3Co_2Fe_{24}O_{41}$ ceramics and of the $\rm
(Ba_{0.5}Sr_{0.5})_3Co_2Fe_{24}O_{41}$ single crystal at selected temperatures.
The single crystal  was  grown and polished with the surface normal
[0001], therefore only $E_{1u}$ modes are active in its
$\textbf{E}^{\omega}\|x,y$ polarized spectra ($\textbf{E}^{\omega}$ denotes the electric vector of the incident radiation). In the spectra of
ceramics, both $A_{2u}$ and $E_{1u}$ modes are IR active, but their
intensities are reduced. We identified only 21\,$E_{1u}$ phonons in
the single crystal and 22 phonons in the ceramics, which is much
less than expected from the factor-group analysis. The discrepancy is
apparently caused by  small intensities and/or
overlapping of some modes. It is worth noting that no new modes appear in the IR
spectra on cooling, i.e.\ no signature of any structural phase transition is seen
below 300\,K. The intensities of the reflection bands only increase
on cooling due to reduced phonon damping with lowering temperature.
The phonon parameters obtained from fits of spectra taken at 10\,K are
listed in Tab.\ I in Suppl. Materials.\cite{Suppl-hexaferrite}

\subsection{THz studies}

 \begin{figure}
           \centering
           \includegraphics[width=82mm]{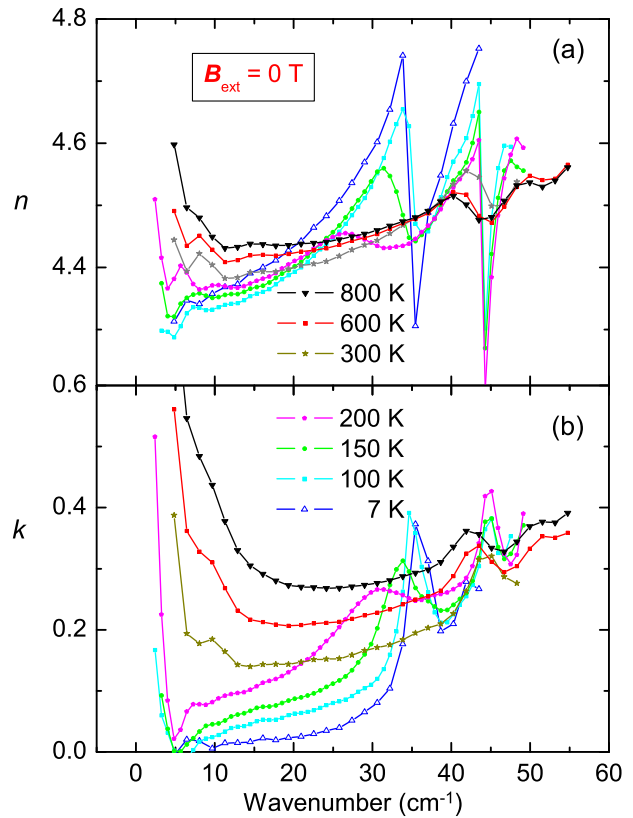}
           \caption{Spectra of the complex refractive index of the
		$\rm(Ba_{0.2}Sr_{0.8})_3Co_2Fe_{24}O_{41}$ ceramics determined
	by THz spectroscopy at various temperatures.}
           \label{fig:ZTHzN7-800}
   \end{figure}

   Fig.~\ref{fig:ZTHzN7-800} shows complex refractive index spectra of the
$\rm(Ba_{0.2}Sr_{0.8})_3Co_2Fe_{24}O_{41}$ ceramics obtained by
time-domain THz spectroscopy between 7 and 800\,K.
 The low-frequency increase in
$n(\omega)$ and $\kappa(\omega)$ occurring above 300\,K is due to
the sample conductivity arising in the microwave region; furthermore, two
clear resonances are seen in the spectra. The higher-frequency one (near $\sim 45\,\cm$)
 is present at all temperatures and it exhibits only a small softening on
heating. The lower-frequency one appears at 250\,K near $25\,\cm$ and
markedly hardens and sharpens on cooling (its frequency reaches 35\,\cm\, at 7\,K).

\begin{figure}
   	  \centering
   	  \includegraphics[width=82mm]{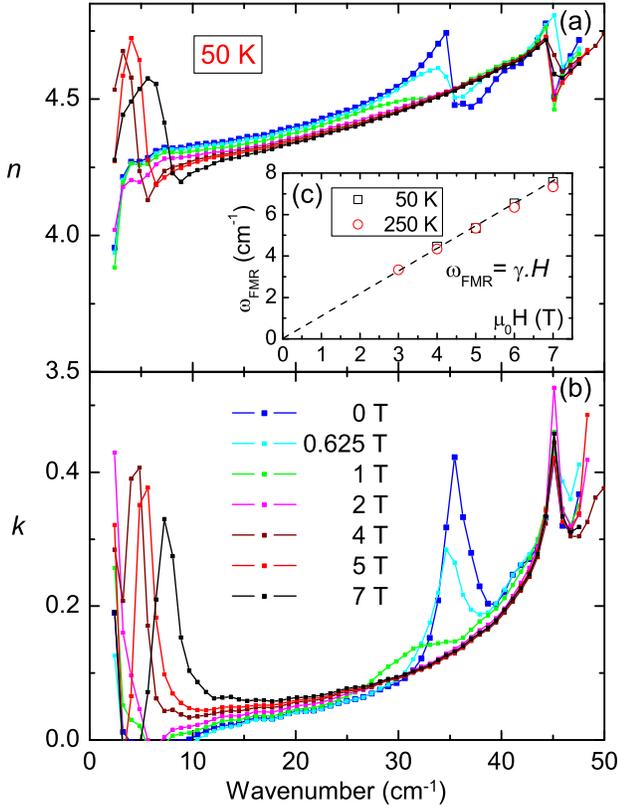}
   	  \caption{Magnetic-field dependence of a) index of refraction and b)
		  extinction coefficient in the THz region measured at
		  50\,K. c)
		  Magnetic field dependence of the ferrimagnetic resonance frequency
		  $\omega_{\rm FMR}$ in
		  $\rm(Ba_{0.2}Sr_{0.8})_3Co_2Fe_{24}O_{41}$ ceramics at 50
		  and 250\,K.}
   	  \label{fig:ZBaTHz50K-H}
     \end{figure}

Upon applying magnetic field, the
lower-frequency resonance broadens, shifts towards lower frequencies and finally
disappears at magnetic field values above 2\,T. This behavior is
shown in Fig.~\ref{fig:ZBaTHz50K-H}
for 50\,K; qualitatively similar magnetic-field dependences were observed
at temperatures
up to 250\,K (see Fig.~3 in  Supplement\cite{Suppl-hexaferrite}). This is a
signature of a magnetic excitation; we assume that it disappears from the
spectra due to the
magnetic phase transition from the transverse conical to a collinear structure.
We expect this mode to remain active up to $T_{\rm con }\cong$ 400\,K when the magnetic structure
changes. Nevertheless, its damping dramatically increases with temperature, so the mode
gradually becomes overdamped. Consequently, above $\sim$
250\,K, it is seen only as a broad featureless
background in the $\kappa(\omega)$ spectra, which overlaps with the conductivity
appearing above 300\,K (see Fig.~3 in  Supplement\cite{Suppl-hexaferrite}). Note that the magnetic mode disappears from the THz
	spectra near the temperature $T_{\rm C3}$ marked in Fig.~\ref{fig:ZBaM-T}.  The
	mode seen near 45\,\cm\,
is apparently a
phonon, because its shape does not change with magnetic field and it remains in the
spectra up to the paramagnetic phase.

As the magnetic field is further increased, another narrow excitation appears
in the low-frequency part of the THz spectra. Its resonance
frequency linearly
increases with the magnetic field as $\omega_{\rm FMR}=\gamma H$ with the
proportionality constant $\gamma=0.032\,\rm THz/T$ (see inset of Fig.~\ref{fig:ZBaTHz50K-H}),
which roughly corresponds to the gyromagnetic ratio of a free
electron ($\gamma=0.028\,\rm
THz/T$). Such behavior is typical of FMRs.\cite{Kittel}
The same resonance is seen up to 250\,K
(the highest value attainable in the magnetic cryostat) and at 7\,T, for all
temperatures, it reaches a frequency of $\approx 7.5$\,\cm. (see
Fig.~\ref{fig:ZBaTHz50K-H}c and Fig.\,3 in Supplement\cite{Suppl-hexaferrite}). Without magnetic field, the FMR can be
observed in the microwave range. These observations are beyond the scope of the
present article and will be presented separately.

\begin{figure}
   	  \centering
   	  \includegraphics[width=82mm]{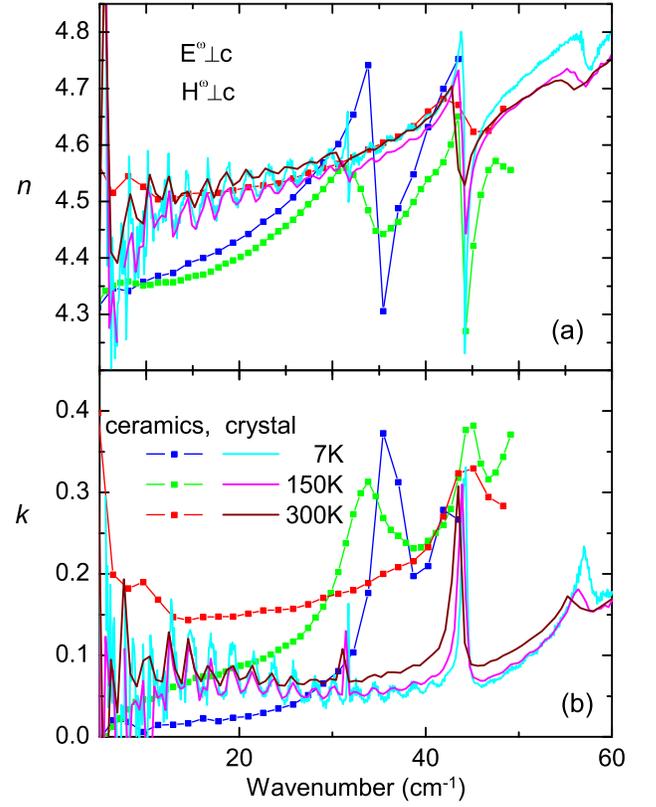}
   	  \caption{Comparison of THz spectra of a) index of refraction and b) extinction
		  coefficient in  $\rm(Ba_{0.2}Sr_{0.8})_3Co_2Fe_{24}O_{41}$
		  ceramics and  $\rm(Ba_{0.5}Sr_{0.5})_3Co_2Fe_{24}O_{41}$
		  single crystal at three temperatures. The spectra of the
		  single crystal are polarized \textbf{E}$^{\omega}\perp z \perp
		  \textbf{H}^{\omega}$, but  the [0001] crystal plane was tilted
		  by 6$^\circ$ from the sample surface, therefore the narrow
		  weak mode near  44\,\cm{} is probably
		  a leakage phonon mode from
		  the \textbf{E}$^{\omega}\parallel z$ polarized spectra.  The
	  oscillations in the spectra of  the single crystal are artifacts due
  to echoes from multiple reflections in the sample. They can be avoided by time
  windowing, but then the spectral resolution is lower and the peak near
  32\,\cm\, is not resolved.}
   	  \label{fig:THz-SC-ceram}
     \end{figure}

In multiferroics,  simultaneously
magnetically and electrically active spin excitation
are called electromagnons. These can be distinguished from magnons by
comparing the polarized IR spectra of crystals taken in all possible
polarizations. Z-hexaferrite single crystals can be
easily grown only in the hexagonal plane, therefore we disposed merely of a (0001)
	single crystal plate. In Fig.~\ref{fig:THz-SC-ceram}, polarized THz spectra of the $\rm (Ba_{0.5}Sr_{0.5})_3Co_2Fe_{24}O_{41}$
single crystal with $\textbf{E}^{\omega} \perp z  \perp\textbf{H}^{\omega}$
(\textbf{H}$^{\omega}$ denoting the magnetic vector of the incident beam)
are compared with the spectra of ceramics. In the single crystal, we
	observed a very weak and narrow excitation at 31\,\cm. Since this mode
	is independent of temperature and external magnetic field (not shown)
	and its frequency is clearly lower than that of the magnetic excitation in
	ceramics, we interpret it as a weak phonon with the $E_{1u}$ symmetry. As
	the spin excitation near 36\,\cm{} is not active in the
	($\textbf{E}^{\omega} \perp z$;
$\textbf{H}^{\omega} \parallel z$) polarized THz spectra\cite{Chun14}, it must be an
electromagnon active for $\textbf{E}^{\omega} \parallel z$.  The existence of this electromagnon was reported already in
2014 at the March APS meeting by Chun\cite{Chun14} but never published.

The two excitations near 44 and 57\,\cm\, are phonons. The former one is
much weaker than the corresponding phonon in ceramics, so it is probably
due to a leakage of an $A_{2u}$($z$) mode.

THz spectra of the $\rm (Ba_{0.5}Sr_{0.5})_3Co_2Fe_{24}O_{41}$ single
crystal measured in external magnetic field  (not shown) revealed  a FMR below 10\,\cm\, with a frequency
identical with that of the FMR in ceramics (see Fig.~\ref{fig:ZBaTHz50K-H}). This provides an evidence that the FMR
has a magnon-like character and that it is active in the $\textbf{E}^{\omega}
\perp z  \perp\textbf{H}^{\omega}$ polarized THz spectra.

THz spectra of $\rm Sr_3Co_2Fe_{24}O_{41}$ ceramics revealed the same
excitations like in $\rm(Ba_{0.2}Sr_{0.8})_3Co_2Fe_{24}O_{41}$, i.e., a phonon
near 45\,\cm\ and an electromagnon near 35\,\cm, (see Fig.~4 in  Supplement\cite{Suppl-hexaferrite}) which vanishes from the spectra  at 250\,K.  This provides an evidence that the
crystalline and magnetic structures of both samples are
almost identical. This was also confirmed by very recent magnetic studies
of $({\rm Ba}_{x}{\rm Sr}_{1-x})_3\rm Co_2Fe_{24}O_{41}$ with $x$ ranging
from 0 to 1.\cite{Tang16}

\subsection{Raman scattering}

In the non-centrosymmetric FE phases, the electromagnons have to be both IR and
Raman active, similarly to the case of BiFeO$_3$.\cite{Skiadopoulou15} In our
case, the electromagnon is active in the \textbf{E}$^{\omega}\parallel
z$-polarized THz spectra, so if the structure is FE (space group $P6mm$),
according to the selection rules in Eq.~(2), the electromagnon has the $A_1$
symmetry and it should be also Raman active in the $z^{2}$-polarized spectra.
	
We measured temperature-dependent Raman spectra of the single crystal
(Fig.~\ref{fig:Raman-sc}) and ceramics (see Fig.~5 in
Supplement\cite{Suppl-hexaferrite}) on cooling down to 4\,K. Indeed,
below 250\,K, a distinctive  excitation appears in the low-frequency part
of the spectra. The inset of Fig.~\ref{fig:Raman-sc} compares
the temperature dependences of frequencies of the Raman-active
excitations in the single crystal and ceramics with those of the IR
active electromagnon in ceramics. The frequencies of all three
excitations are very similar and all of them decrease on heating towards
$T_{\rm con}$. Nevertheless, a detailed analysis reveals some
differences: (i) The Raman-active mode seen in the ceramics has
a frequency systematically higher by 7--10\,\cm than the IR-active
one observed in the same sample. (ii) Upon heating, the Raman-active mode detected in
the single crystal softens faster than those in ceramics.
(iii) In both samples, the Raman-active spin excitation has
a damping systematically 2--3 times higher than the IR-active mode.

The first difference can be possibly a consequence of an angular dependence of the
mode frequency (oblique mode) relevant to the single-crystalline ceramic
grains  (in fact, this Raman mode was observed only in few
grains, and we suppose that their  $z$ crystal axes were
oriented almost parallel to  the sample
surface). The second observation could be theoretically owing to somewhat
different Ba concentrations in the ceramics ($x=0.2$) and the single crystal
($x=0.5$), but since we know that the magnetic structure is identical for both
compositions, this possibility is not likely. Instead, we note that
owing to the sintering, there may be a varying mechanical stress, which may
influence the temperature dependences of the mode frequencies in the individual
grains. Finally, the different dampings of the modes in Raman and THz spectra
do not support the idea of identical modes either.
Consequently, we are not sure that the spin excitations seen in both THz
and Raman spectra are the same. In case the Raman mode is not identical with the
IR-active one, the Raman mode could be due to an antiferromagnetic
resonance or more probably multi-magnon scattering involving magnons with
the highest magnon density of states (DOS) .

\begin{figure}
   	  \centering
   \includegraphics[width=85mm]{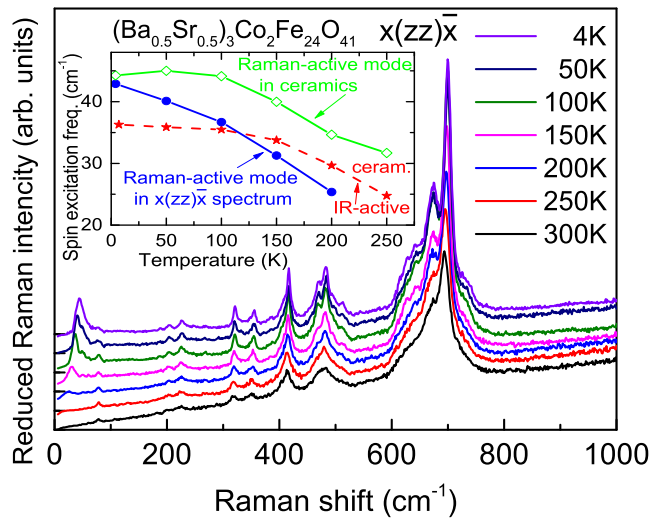}
   	  \caption{Temperature dependence of $x(zz)\overline{x}$ Raman
	  spectra of the $\rm(Ba_{0.5}Sr_{0.5})_3Co_2Fe_{24}O_{41}$ single
	  crystal. The inset shows the temperature dependence of spin excitation
	  frequencies obtained from these spectra and from unpolarized Raman and THz spectra of the
	  $\rm(Ba_{0.2}Sr_{0.8})_3Co_2Fe_{24}O_{41}$ ceramics.}
	  \label{fig:Raman-sc}
     \end{figure}

In the opposite case, if the Raman and IR-active excitations
corresponded to the same vibrational mode, it would be an argument
	supporting the FE $P6mm$ crystal symmetry of the
	Z-hexaferrite. Then,  the FE polarization \textbf{P} would be oriented along the $z$ axis, which is in
	contradiction with Ref.~\onlinecite{Chun12}, where magnetic-field
		induced changes in polarization in the
		hexagonal plane were observed. One might also hypothetically
		assume that at zero magnetic field,
		$\textbf{P} \parallel z$, and that
		\textbf{P} tilts away from the $z$
		axis in external magnetic field; in such a case, the crystal
		structure would change to monoclinic. However, in the FE
		phase, a much higher number of phonons than observed should be
		present in both IR and Raman spectra. For example, at 4\,K, we resolved 20 modes in
			$z^{2}$-polarized Raman spectra and 21 IR-active phonons
			in \textbf{E}$^{\omega}\perp z$ spectra. These
			numbers are in better agreement with the 26
			$A_{1g}(z^2)$ and $36E_{1u}(x,y)$ modes predicted
			for the paraelectric $P6_{3}/mmc$ phase than with the
			54 $A_{1}$($z^2$) and 70 $E_{1}$($x,y$) modes expected for the FE
			$P6mm$ symmetry.
		
We can conclude that our spectra do not support any FE distortion in zero
magnetic field. This is in agreement with the known literature;
up to now, in the structural studies of Z-hexaferrite
$({\rm Ba}_{x}{\rm Sr}_{1-x})_3\rm Co_2Fe_{24}O_{41}$, no polar space group was
resolved\cite{Takada03,Takada06} and the electric polarization $\textbf{P}
\perp z$ was observed only in an external magnetic
field.\cite{Kitagawa10,Chun12} Nevertheless, for final proving or
disproving of the polar crystal structure, we propose further
complementary experiments, such as second-harmonic generation or
high-resolution electron diffraction.

\subsection{Inelastic neutron scattering}

\begin{figure}[t]
	  \centering
	  \includegraphics[width=\columnwidth]{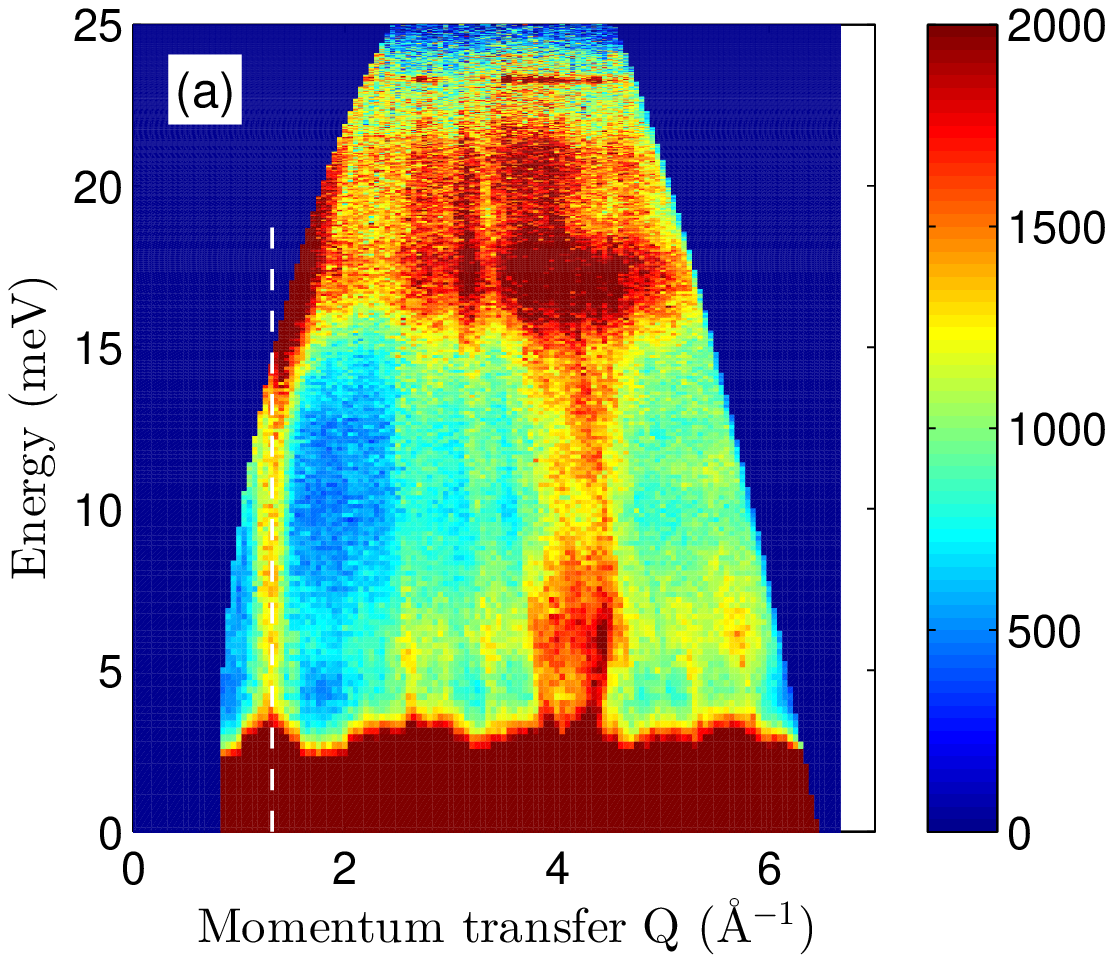}
      \includegraphics[width=85mm]{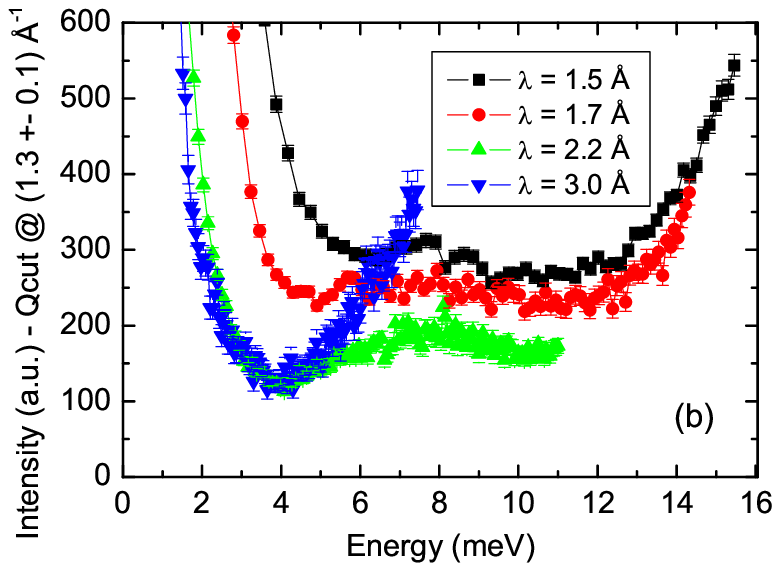}%
%	  \begin{picture}(0, 0)(0,0)
%		  \setlength{\unitlength}{1mm}
%		  \put(-65,7){\vector(0,1){12}}
%	  \end{picture}
	  \caption{(a) INS intensity as a function of momentum and energy
	  transfers, measured on $\rm(Ba_{0.2}Sr_{0.8})_3Co_2Fe_{24}O_{41}$
	   at 5\,K. The dashed line marks the signal corresponding to the magnon
	   branch. The remaining signal comes from phonon DOS. The
	  energy resolution was $\Delta E=1.5\,\rm meV$, the wavelength of the
	  incident neutrons was 1.5\AA. (b) Magnon DOS calculated from the
	  INS spectra for $1.2\leq Q \leq1.4\,\mbox{\AA}$ taken with various  wavelengths $\lambda$ of the incident neutrons. }
	  \label{fig:INS}
\end{figure}

Some electromagnons activated in the THz dielectric spectra by exchange coupling
 have wavevectors from the Brillouin-zone
boundary,\cite{Stenberg09,Stenberg12,Kadlec13} where the magnon DOS attains a maximum.
Such magnons should be recognized by the corresponding maxima of
intensity in the INS spectra. Using the powder, we have performed INS experiments with
various energy resolutions. Fig.~\ref{fig:INS}a shows a map representing the
orientation-averaged scattering intensity at $T=5\,\rm K$. The high
INS intensity seen as a column at $Q \approx 4\,\rm \AA^{-1}$ and for all $Q$ at
energies above 15 meV corresponds to the phonon DOS.
Near the neutron momentum transfer value of $Q=1.3\,\rm\AA^{-1}$,  a magnetically active
branch (marked by the dashed line) extends in energy up to at least 20\,meV. The
magnon DOS  (proportional to the scattering intensity  integrated over
the interval $1.2\leq Q \leq1.4\,\mbox{\AA}$) exhibits a small maximum
near $8\,\mbox{meV} \approx 64\,\cm$ (see Fig.~\ref{fig:INS}b). Near this
energy we see only a weak excitation in the \textbf{E}$^{\omega}\perp z$
polarized IR reflectivity spectra (see Fig.~\ref{fig:IRrefl} and Tab.\ I in
Supplement\cite{Suppl-hexaferrite}); based solely on our data, we cannot
decide whether this is an electromagnon or a phonon. The
electromagnon  energy, according to our THz  spectra, amounts to
4--5.5 meV, but at the low temperatures used in INS experiments, at
this energy transfer value, a minimum in magnon DOS is seen
(Fig.~\ref{fig:INS}b). Subsequently, the wavevector of the electromagnon
is most probably not from the Brillouin zone center or boundary. This is
	rather surprising because recent polarized INS studies of a Y-type
	hexaferrite claimed that the observed electromagnon with similar features as the one in Z-hexaferrite was a zone-center mode.\cite{Nakajima16} .

In the transverse-conical structure, the
electromagnon  can induce an
oscillating electric polarization along the $z$ axis. Therefore we suggest that it
is activated by the exchange-striction mechanism. Note that in both
Z- and Y-type hexaferrites, the static polarization
\textbf{P} appears due to the inverse Dzyaloshinskii-Moriya
interaction \cite{Kitagawa10,Tokura14-review} perpendicularly to
the $z$ axis, if one applies an external  magnetic field
$\textbf{H} \perp z$ whereby the transverse conical structure is stabilized.\cite{Chun12,Nakajima16} However, the magnetic
structure of the $Y$-type hexaferrite is longitudinally conical at $\mu_{0}H =
0\,T$,\cite{Kida09a,Kida11} whereas that of the $Z$-hexaferrite at zero and small magnetic fields is transverse conical. The electromagnons
in both materials share several features: (i) they are active in \textbf{E}$^{\omega}\parallel z$ even in zero
magnetic field, (ii) their frequencies shift with external magnetic field and, finally, (iii) they
disappear from the spectra above some threshold magnetic field, when the transverse or longitudinal conical magnetic structures disappear.

\section{Conclusion}
In conclusion, by using several complementary spectroscopic techniques, we
have obtained a comprehensive set of information on the spin and lattice
dynamics of the Z-type hexaferrite compounds $({\rm Ba}_{x}{\rm Sr}_{1-x})_3\rm
Co_2Fe_{24}O_{41}$, in broad frequency and temperature ranges.

In the low-temperature THz spectra of $({\rm Ba}_{x}{\rm Sr}_{1-x})_3\rm
Co_2Fe_{24}O_{41}$ ceramics, a soft spin excitation near 35\,\cm\  was
discovered, whose frequency softens  on heating towards
$T_{\rm con}\approx$ 400\,K  and its damping increases. An external magnetic field exceeding 2\,T induces a change of magnetic structure and the spin excitation vanishes from the
THz spectra. THz spectra obtained on a single
crystal revealed the same magnon in the \textbf{E}$^{\omega}\parallel z$ polarized spectra, therefore we claim that it is an
electromagnon. Since the excitation is observed in the transverse conical magnetic structure, we propose that it is activated by the exchange striction mechanism.

A spin excitation with a similar frequency was discovered in Raman spectra. Should it be the same
electromagnon, the sample would be FE with a polarization
$\textbf{P} \parallel z$ and the $P6mm$ space group. This is rather
unlikely, because up to now, only $\textbf{P} \perp z$ oriented polarization was
observed  and the numbers of phonon modes observed in the IR and
Raman spectra are much lower than those allowed in the FE phase.
Nevertheless, further structural, magnetoelectric and
second-harmonic-generation experiments
appear necessary to clearly prove or disprove a polar phase in $({\rm Ba}_{x}{\rm Sr}_{1-x})_3\rm Co_2Fe_{24}O_{41}$.

Upon applying magnetic field higher than 3\,T,  in the low-frequency part of the THz
spectra, a narrow excitation appears whose frequency linearly increases with magnetic field. Its behavior is independent on temperature (investigated up to 250\,K) and since the proportionality constant of the resonance frequency on the magnetic field corresponds to the gyromagnetic ratio of a free electron, we interpret this excitation as the ferromagnetic resonance.

\begin{acknowledgments}
This work was supported by the Czech Science Foundation projects 15-08389S and
14-18392S, the program of Czech Research Infrastructures, project LM2011025 and
M\v{S}MT project LD15014. The experiments in ILL Grenoble were carried out
within the project LG14037 financed by the Ministry of Education of the Czech Republic.

\end{acknowledgments}

\newpage
\section{Electromagnon  in Z-type hexaferrite $({\rm Ba}_{x}{\rm Sr}_{1-x})_3\rm Co_2Fe_{24}O_{41}$ - Supplementary materials}

%\vspace*{-8pt}
           \parbox{\columnwidth}{ \includegraphics[width=82mm]{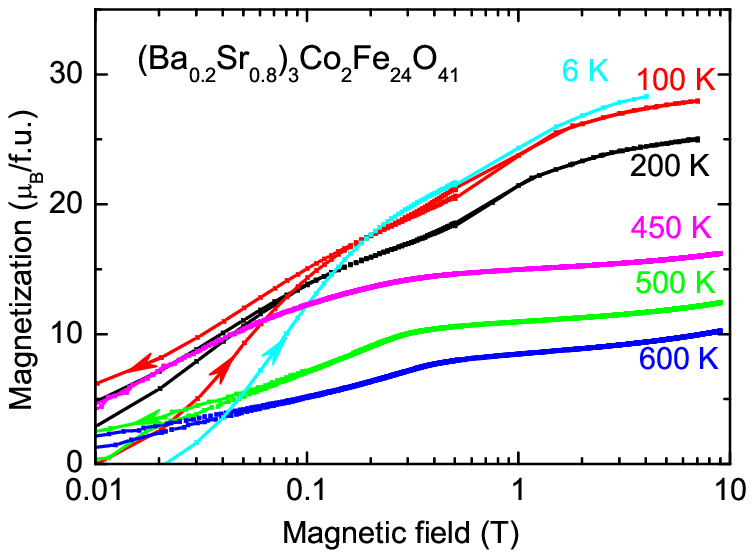}
            {\noindent \small FIG. 1: Magnetization curves of polycrystalline
            $\rm(Ba_{0.2}Sr_{0.8})_3Co_2Fe_{24}O_{41}$ taken at various temperatures. }}
	    \par\vspace{14mm}
   	\parbox{\columnwidth}{\protect\vphantom{,}\includegraphics[width=82mm]{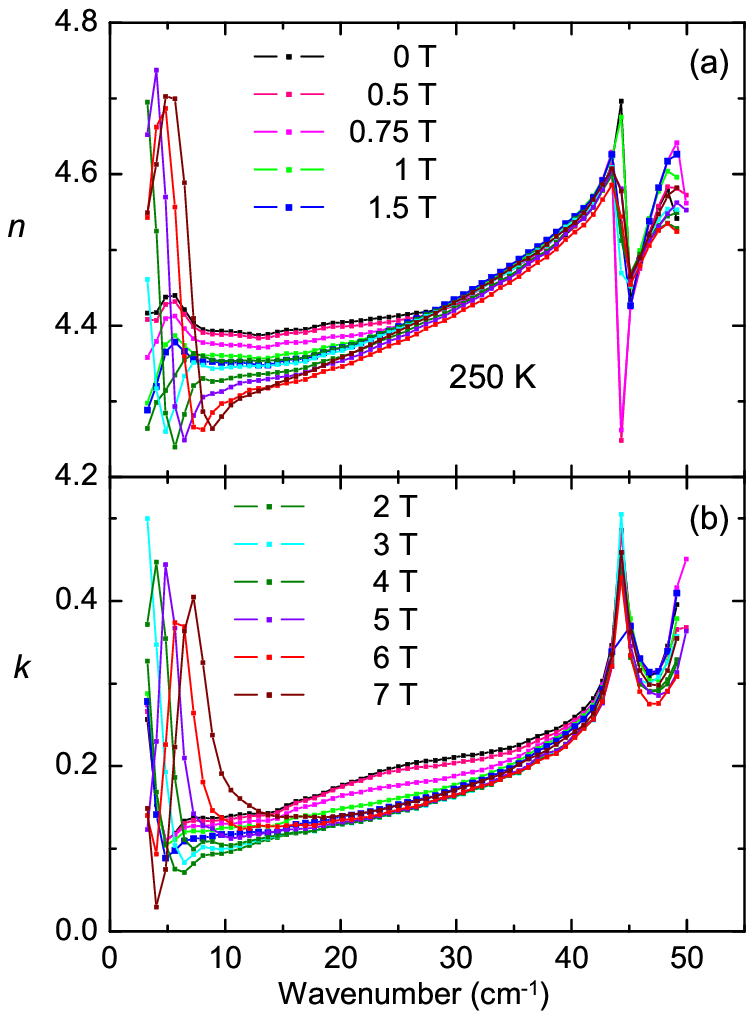}
   	  {\noindent \small FIG. 3: Magnetic-field dependence of a) index of refraction and (b)
	  extinction coefficient of $\rm(Ba_{0.2}Sr_{0.8})_3Co_2Fe_{24}O_{41}$ ceramics measured at 250\,K. }}

	    \parbox{\columnwidth}{\includegraphics[width=82mm]{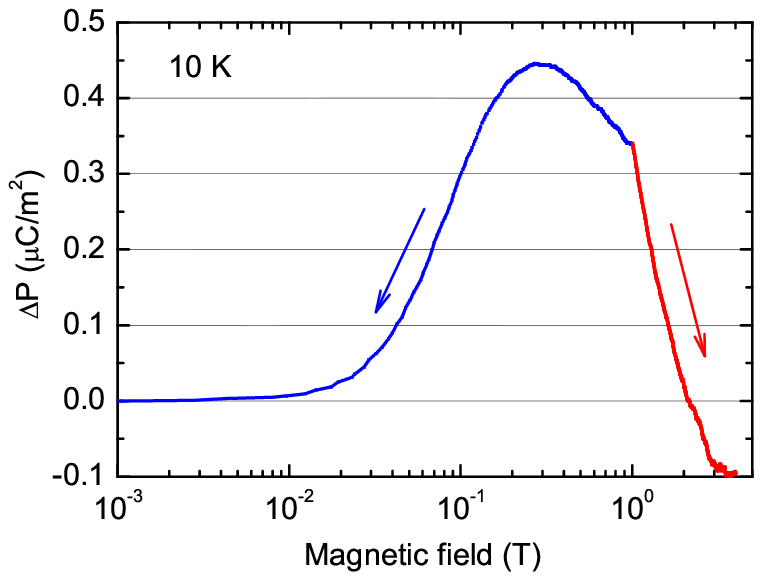}
	{\noindent \small FIG. 2: Magnetic-field dependence of polarization changes
	at $T=10\,\rm K$ for
		the $\rm(Ba_{0.2}Sr_{0.8})_3Co_2Fe_{24}O_{41}$ ceramics. The blue
		and red curves show changes upon decreasing and
		increasing magnetic field, respectively.}}
	    \par\vspace*{7mm}
	  \parbox{\columnwidth}{\includegraphics[width=82mm]{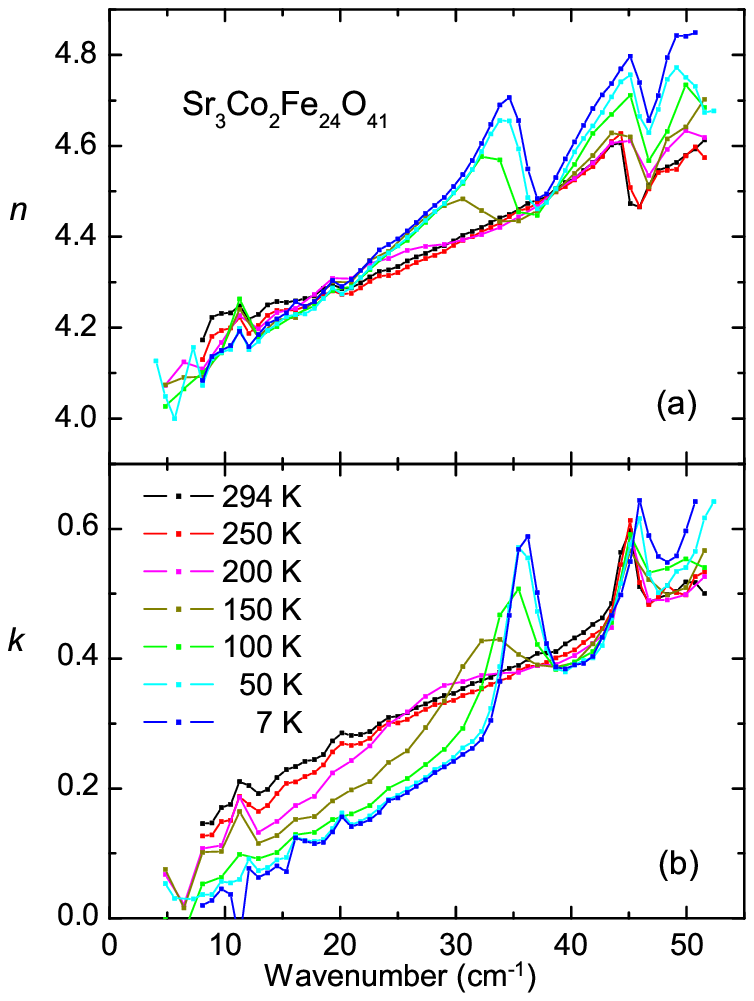}
	      {\noindent \small FIG. 4: Temperature dependence of THz spectra of the
		      complex refractive index of $\rm Sr_3Co_2Fe_{24}O_{41}$ ceramics}}

		      \addtocounter{figure}{4}
 \begin{figure*}
	    %\parbox{\columnwidth}
	    { \includegraphics[width=82mm]{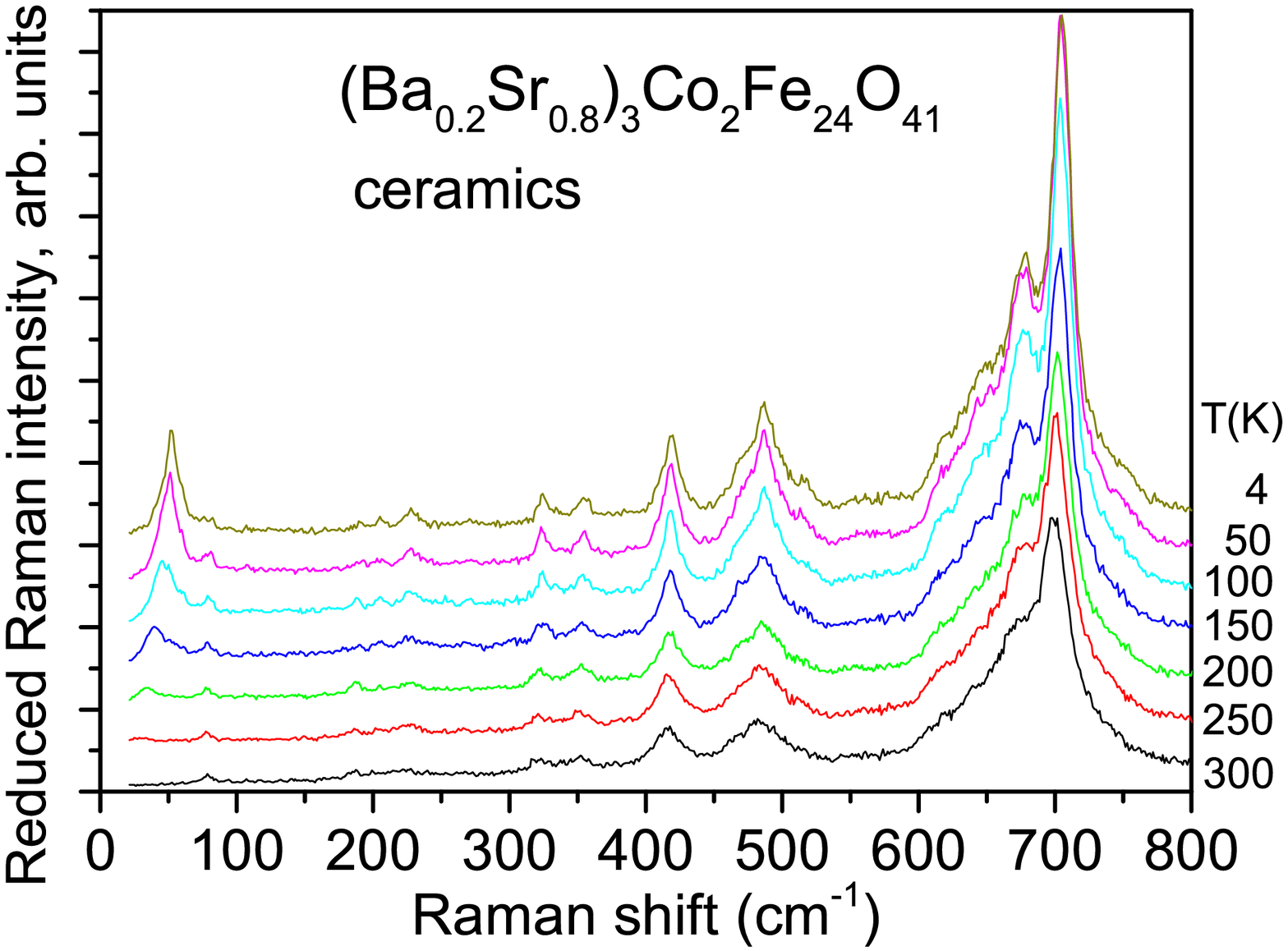}
   	  {\noindent \\
   \small FIG. 5: Temperature dependence of Raman spectra of the
		  $\rm(Ba_{0.2}Sr_{0.8})_3Co_2Fe_{24}O_{41}$ ceramics measured
		  with the polarizer and analyzer set parallel. It is
		  important to stress that the electromagnon seen below 50\,\cm\, was
  observed only in few ceramic grains. This is consistent with its activity only
  in the $z^2$ Raman spectra. The grains are randomly oriented and only few of
  them have their $z$ axes lying in the sample plane.}}
    \end{figure*}

\nopagebreak
\begin{table*}
  \small
\caption{Comparison of phonon fitting parameters
describing IR reflectivity and THz transmission spectra of (Ba$_{0.2}$Sr$_{0.8}$)$_3$Co$_2$Fe$_{24}$O$_{41}$ ceramics
and those of the (Ba$_{0.5}$Sr$_{0.5}$)$_3$Co$_2$Fe$_{24}$O$_{41}$ single
crystal (\textbf{E}$^{\omega}\perp z$ and \textbf{H}$^{\omega}\perp z$) obtained
at $T=10\,\rm K$. The high-frequency contributions to permittivity were
$\varepsilon_{\infty} = 4.92$ in the ceramics and $\varepsilon_{\infty} = 4.79$
in the single crystal.
The phonons seen in the IR reflectivity spectra of the single crystal are assumed to
have the $E_{2u}$ symmetry. The two phonons in THz transmission
spectra below 60\,\cm are much weaker in the crystal than in the ceramics,
therefore we assume they have the $A_{2u}$ symmetry and they
appear  as a leakage from the \textbf{E}$^{\omega}\parallel
z$ spectra (the \textit{z} crystal axis was tilted by 6$^\circ$ from
the normal crystal plane). The oscillators present only in the
ceramics are assumed to have the $A_{2u}$ symmetry.}

\begin{ruledtabular}
	\bgroup
\begin{tabular}{lcccccc}
  & \multicolumn{3}{c}{Ceramics} & \multicolumn{3}{c}{Single crystal} \\
\cline{2-4} \cline{5-7}
   Symmetry  & $\omega_{\rm TO}$ (\cm) & $\gamma_{\rm TO}$ (\cm) & $\Delta\varepsilon$  & $\omega_{\rm TO}$ (\cm) & $\gamma_{\rm TO}$ (\cm) &  $\Delta\varepsilon$ \\
\hline

\hline
 A$_{2u}$? & 45.26 & 1.69 & 0.12 & 43.78  & 0.24  & 0.02 \\[-3pt]
 A$_{2u}$? &  &  &  & 56.91  & 1.92  & 0.04                    \\
 E$_{2u}$ &  &  & & 66.33  & 9.19  & 0.24                    \\
  & 87.09  & 29.80 & 4.03 &   &   &                    \\
 E$_{2u}$ &  &   &     & 98.43 & 19.29 &  1.80                \\
 E$_{2u}$ &  &  &    & 109.56 & 10.83  & 0.82                 \\
 A$_{2u}$  & 123.04 & 8.29 & 0.33 &   &   &                                 \\
 A$_{2u}$  & 138.23 & 8.65 & 0.15 &   &   &                                 \\
 E$_{2u}$  & 160.06 & 15.62 & 0.18  & 166.91 & 8.79  & 0.24            \\
 E$_{2u}$  & 172.04 & 11.37 & 0.17  & 176.32 & 9.32 & 0.21             \\
 A$_{2u}$  & 190.02 & 17.04 & 0.36 &   &   &                                       \\
 E$_{2u}$  & 211.84 & 14.00 & 0.46  & 209.70 & 28.11 & 0.68            \\
 A$_{2u}$  & 222.54 & 7.28 & 0.14 &   &   &                                        \\
E$_{2u}$   & 240.52 & 12.85 & 0.54  & 239.23 & 11.88   & 1.70         \\
A$_{2u}$   & 252.93 & 11.33 & 0.30 &   &  &                                       \\
 &   &   &    & 267.91 & 29.74  & 0.89                                                 \\
E$_{2u}$   & 285.88 & 13.01 & 0.77  & 285.02 & 8.46 & 2.01            \\
E$_{2u}$   & 298.72 & 16.58 & 1.31 & 297.01 & 11.34  & 0.89          \\
E$_{2u}$   & 309.85  & 6.11 & 0.10 & 309.85 & 12.21    & 1.22         \\
E$_{2u}$   & 320.97 & 17.47 & 1.30  & 323.54 & 4.62  & 1.01            \\
E$_{2u}$   & 359.49 & 6.46 & 0.05  & 358.83 & 11.09 & 0.22            \\
E$_{2u}$  &\multirow{2}{*}{373.18} & \multirow{2}{*}{42.74} & \multirow{2}{*}{1.53}   & 374.90 & 15.91  & 0.59     \\
E$_{2u}$  &   &  &    & 385.81 & 23.89  & 0.50                                  \\
E$_{2u}$   & 445.08 & 39.37 & 0.59  & 431.25 & 9.52  & 1.04          \\
A$_{2u}$   & 503.29 & 20.03 & 0.08 &   &   &                                    \\
E$_{2u}$   & 540.95& 12.12 & 0.02  & 539.23 & 27.63   & 0.79        \\
E$_{2u}$  & 566.62 & 59.22 & 0.81  & 560.49 & 21.00 & 0.64      \\
E$_{2u}$   & 595.73 & 36.54 & 0.33  & 590.59 & 14.14  & 0.38       \\
            \end{tabular}
	    \egroup
 \end{ruledtabular}
\end{table*}

\end{document}